# Nonlinear waves in stratified Taylor–Couette flow. Part 2. Buoyancy flux.

## Colin Leclercq[1]†, Jamie L. Partridge[2], Colm-Cille P. Caulfield[2], Stuart B. Dalziel[2] and Paul F. Linden[2]

[1]School of Mathematics, University of Bristol, University Walk, Bristol BS8 1TW, UK

[2]Department of Applied Mathematics and Theoretical Physics, Centre for Mathematical Sciences, Wilberforce Road, Cambridge CB3 0WA, UK

This paper is the second part of a two-fold study of mixing, i.e. the formation of layers and upwelling of buoyancy, in axially stratified Taylor–Couette flow, with fixed outer cylinder. In a first paper, we showed that the dynamics of the flow was dominated by coherent structures made of a superposition of nonlinear waves. (Mixed)-ribbons and (mixed)-cross-spirals are generated by interactions between a pair of linearly unstable helical modes of opposite 'handedness', and appear to be responsible for the formation of well-mixed layers and sharp density interfaces. In this paper, we show that these structures are also fully accountable for the upwards buoyancy flux in the simulations. The mechanism by which this occurs is a positive coupling between the density and vertical velocity components of the most energetic waves. This coupling is primarily caused by diffusion of density at low Schmidt number $Sc$, but can also be a nonlinear effect at larger $Sc$. Turbulence was found to contribute negatively to the buoyancy flux at $Sc = 1, 10, 16$, which lead to the conclusion that mass upwelling is a consequence of chaotic advection, even at large Reynolds number. Artificially isolating the coherent structure therefore leads to excellent estimates of the flux Richardson numbers $Ri_f$ from the DNS. We also used the theoretical framework of Winters *et al.* (1995) to analyse the energetics of mixing in an open control volume, shedding light on the influence of end effects in the potential energy budget. The potential connection with the buoyancy flux measurements made in the recent experiment of Oglethorpe *et al.* (2013) is also discussed.

**Key words:**

## 1. Introduction

In this paper, we follow up on the investigation of the dynamics of mixing in axially stratified Taylor–Couette (STC) flow, revisiting Oglethorpe *et al.* (2013)'s experiments with linear instability analysis and direct numerical simulations. The first part of this endeavour (Leclercq *et al.* (????b), hereafter referred to as Pt. 1) was focussed on the formation of non-axisymmetric patterns at large Reynolds number, previously unreported by Oglethorpe *et al.* (2013), but potentially responsible for the creation of well-mixed density layers and sharp density interfaces in their experiments. In this second contribution, we concentrate on the energetics of mixing and establish a firm link between the coherent structures identified in Pt. 1, and the upwelling of buoyancy. This study describes a physical mechanism based on the presence of nonlinear waves, achieving large mixing efficiency in the weakly turbulent regime, for low Schmidt $Sc$ numbers (ratio between momentum and density diffusivities).

† Email address for correspondence: c.leclercq@bristol.ac.uk

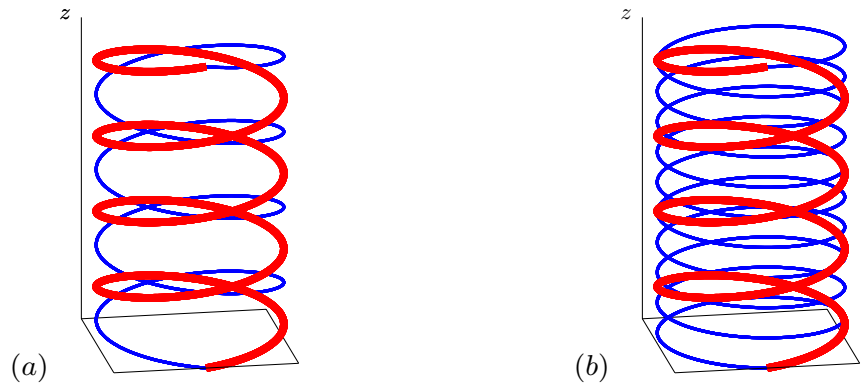

Figure 1. Schematics of pair of dominant helical modes in a (*a*) ribbon and a (*b*) mixed-ribbon structures.

Understanding mass transport in stratified shear flows is important in the context of geophysical fluid dynamics. Indeed, the mechanisms of dyapicnal mixing which are responsible for the upwelling of abyssal waters to the upper layers of the ocean remain still largely unknown. Yet, this ingredient is essential to the closure of the meridional overturning circulation (Wunsch & Ferrari 2004), a key process in all climate models. Besides, from a more fundamental point of view, appropriate quantification of the mixing efficiency, i.e. the portion of mechanical energy expended for irreversible mixing (Caulfield & Peltier 2000) as opposed to stirring, remains a challenge today in the context of stratified turbulence (Ivey *et al.* 2008; Davies Wykes *et al.* 2015). Despite the potentially limited relevance of a wall-bounded flow like STC to the context of geophysical fluid dynamics, it is an amenable model of stratified shear flow (with stratification perpendicular to shear) which allows us to address the two fundamental questions of mass transport and mixing efficiency both numerically and experimentally.

In Pt. 1, we identified coherent structures emerging from the interaction of a pair of linearly unstable helical modes of opposite 'handedness'. These structures, represented schematically in figure 1, may be called *ribbon* (Demay & Iooss 1984) if the waves are mirror-symmetric and have equal amplitude, *mixed-ribbon* (Altmeyer & Hoffman 2014) if the pitch of the helices are not exactly opposite but the amplitudes are still comparable, and *(mixed-)cross-spirals* (Altmeyer & Hoffman 2014) if the amplitudes are significantly different (see figure 1). Whereas ribbon structures have been known for a long time in the unstratified, counter-rotating case (Demay & Iooss 1984; Tagg *et al.* 1989), they had apparently never been observed at such high Reynolds numbers $Re = O(10^3 - 10^4)$ before. Taking advantage of our axially periodic model of STC, we are able to perform a spectral decomposition of the buoyancy flux term, revealing the dominant contribution of these structures to the global upwelling of mass.

Using the theoretical framework set out by Winters *et al.* (1995), we also derive kinetic and potential energy budgets for our axially periodic flow. In order to investigate mixing, we break down potential energy into its available and background contributions, thereby separating isopycnal from diapycnal transport. We discuss the advantages and limits of the periodic model in diagnosing mixing. In particular, this analysis sheds light on the specific role played by end-plates in real experiments. We also discuss the relevance of the buoyancy Reynolds number, a widely used parameter quantifying the strength of turbulence in the context of stably stratified flows, with respect to the mixing mechanism identified in this article.

The plan of the paper is as follows. In §2, we introduce the governing equations and numerical methods. In §3, we present our analysis of the energetics of mixing in a vertically periodic domain and show the dominant role of nonlinear waves in the mass transport mechanism. Finally, in §4, we discuss the relevance of turbulence to this surprisingly efficient mixing mechanism. In particular, we revisit the universal flux law of Oglethorpe *et al.* (2013) and discuss the potential relevance of our low $Sc$ findings to their high $Sc$ experimental results. We conclude in §5.

## 2. Problem formulation

The problem formulation is the same as in Pt. 1, regarding both the governing equations and numerical methods. We will briefly introduce the model here in order to provide all the control parameters and notations, but the reader is referred to Pt. 1 for more details.

The geometry is characterised by two nondimensional parameters, the radius ratio $\eta$ and aspect ratio $\Gamma$, defined as

$$\eta := \frac{r_i}{r_o} \quad \text{and} \quad \Gamma := \frac{h}{\Delta r}, \tag{2.1}$$

where $\Delta r := r_o - r_i$ is the gap between the outer and inner radii, $r_o$ and $r_i$, and $h$ is the 'apparent height' of our axially-periodic domain. The Reynolds number is defined as

$$Re := \frac{r_i \Omega \Delta r}{\nu}, \tag{2.2}$$

where $\Omega$ is the rotation rate of the inner cylinder and $\nu$ is the kinematic viscosity. The density field is decomposed as

$$\rho_{\text{tot}} = \rho_0 + \underbrace{\bar{\rho} + \rho}_{\tilde{\rho}}, \tag{2.3}$$

where $\rho_0$ is a reference density and $\tilde{\rho}$ a deviation made of a linear background stratification of buoyancy frequency $N$ ($g$ denotes gravity),

$$\bar{\rho} = -\frac{\rho_0 N^2}{g} z, \tag{2.4}$$

and a periodic perturbation $\rho$ in both the azimuthal and axial directions. Stratification is characterised by a bulk Richardson number

$$Ri := \frac{N^2}{\Omega^2}, \tag{2.5}$$

and the Schmidt number

$$Sc := \frac{\nu}{\kappa} \tag{2.6}$$

based on the ratio between the diffusivity of mass $\kappa$ and kinematic viscosity.

The flow is governed by the incompressible Navier–Stokes equations in the Boussinesq approximation, with no need to include 'centrifugal buoyancy' here (cf. Pt. 1). We also note that the Boussinesq approximation inevitably breaks for infinitely long cylinders, but it may be considered valid over the (finite) vertically periodic domain. Using typical scales $\Delta r$ for length, $r_i \Omega_i$ for velocity and

$$\Delta \rho := \bar{\rho}(z - h/2) - \bar{\rho}(z + h/2) = \frac{\rho_0 N^2 h}{g} \quad (2.7)$$

for density, the nondimensional governing equations for the velocity field $\mathbf{u}$, expressed as $\mathbf{u} = u\mathbf{e}_r + v\mathbf{e}_\theta + w\mathbf{e}_z$ in cylindrical coordinates, and the perturbation density $\rho$ read

$$\partial_t \mathbf{u} + \mathbf{u} \cdot \nabla \mathbf{u} = -\nabla p + \frac{1}{Re}\nabla^2 \mathbf{u} - \beta\rho\mathbf{e}_z, \quad (2.8)$$

$$\partial_t \rho + \mathbf{u} \cdot \nabla \rho - \frac{w}{\Gamma} = \frac{1}{Re\,Sc}\nabla^2 \rho, \quad (2.9)$$

$$\nabla \cdot \mathbf{u} = 0, \quad (2.10)$$

where $p$ is a potential based on the actual pressure and finally

$$\beta := \Gamma \frac{(1-\eta)^2}{\eta^2} Ri \quad (2.11)$$

is the nondimensional version of the reduced gravity $g\Delta\rho/\rho_0$. The boundary conditions are impermeability and no-slip for the velocity, and no-flux $\partial_r \rho = 0$ for density (modelling salt in water). The parameter values of our simulations are the same as in Part 1, and are recalled in table 1, together with some useful output metrics which will be discussed throughout the paper.

Direct numerical simulation were performed using an adapted version of Shi *et al.* (2015)'s pseudospectral code, based on a representation of the fields $\mathbf{q} := (\mathbf{u}, p, \rho)$ of the form

$$\mathbf{q}(r, \theta, z, t) = \sum_{m=-n_\theta/2}^{n_\theta/2} \sum_{k=-n_z/2}^{n_z/2} \mathbf{q}_{m,k}(r,t) \exp[\mathrm{i}(m\theta + kk_0 z)], \quad (2.12)$$

where $k_0 := 2\pi/\Gamma$ and $\mathbf{q}_{-m,-k} = \mathbf{q}^*_{m,k}$ ($^*$ denotes the complex conjugate). For more details about implementation, the reader is referred to section 2 in Pt. 1. For more details about linear stability methods, the reader is referred to Leclercq *et al.* (????a).

## 3. The energetics of mixing

### 3.1. *Kinetic and potential energy budgets for a vertically periodic system*

Since we are considering a vertically periodic model of STC, we start section §3 by deriving kinetic and potential energy budgets in an open control volume. In the following equations, energies (resp. powers) are made nondimensional with respect to $\rho_0(r_i\Omega)^2\Delta r^3$ (resp. $\rho_0(r_i\Omega)^3\Delta r^2$). As usual in stratified shear flows, the kinetic energy

$$E^k := \frac{1}{2}\iiint_{\mathcal{V}} \mathbf{u} \cdot \mathbf{u} \, \mathrm{d}\mathcal{V} \quad (3.1)$$

budget involves three terms

$$\frac{\mathrm{d}E^k}{\mathrm{d}t} = \mathcal{P} - \epsilon - \mathcal{B}, \quad (3.2)$$

where

$$\mathcal{P} := r_i \iint_{r=r_i} \tau_{r\theta} \, \mathrm{d}\theta \, \mathrm{d}z, \quad \epsilon := \iiint_{\mathcal{V}} \tau_{ij}\partial_j u_i \, \mathrm{d}\mathcal{V} \quad \text{and} \quad \mathcal{B} := \beta \iiint_{\mathcal{V}} \rho w \, \mathrm{d}\mathcal{V} \quad (3.3)$$

are respectively a production term due to the torque at the inner boundary, dissipation and the buoyancy flux (we note that $-\mathcal{B}$ is sometimes used for the definition of the buoyancy flux, cf. Winters *et al.* (1995); Peltier & Caulfield (2003)). In these definitions, $\tau_{ij}$ denotes the viscous stress tensor, expressed as

$$\tau_{ij} = \frac{1}{Re}(\partial_i u_j + \partial_j u_i) \qquad (3.4)$$

in index notation for a Newtonian fluid.

|   | A1 | A2 | B | C | D | E | F | G | S | CT |
|---|---|---|---|---|---|---|---|---|---|---|
| $Re$ | 2000 | 5000 | 5000 | 5000 | 10000 | 5000 | 5000 | 245 | 816 |
| $Ri$ | 3 | 2 | 3 | 10 | 3 | 3 | 3 | 4.15 | 0.373 |
| $Sc$ | 1 | 1 | 1 | 1 | 1 | 10 | 1 | 16 | 16 |
| $\eta$ | 0.417 | 0.417 | 0.417 | 0.417 | 0.417 | 0.417 | 0.625 | 0.769 | 0.769 |
| $\Gamma$ | 3 | 3 | 3 | 3 | 3 | 3/8 | 3 | 3.25 | 3.25 |
| $Ri_f(\%)$ | 19 | 17 | 15 | 16 | 22 | 15 | 6.0 | 11 | 2.3 | 2.8 |
| $\mathcal{P}$ | 4.5 | 4.2 | 8.1 | 7.9 | 7.3 | 12 | 8.8 | 9.9 | 1.6 | 3.3 |
| $\mathcal{B}$ | 0.85 | 0.73 | 1.2 | 1.3 | 1.6 | 1.8 | 0.53 | 1.1 | 0.037 | 0.092 |
| $Re_b$ | 1.3 | 1.2 | 3.5 | 2.2 | 0.58 | 3.6 | 2.8 | 12 | 5.4 | 123 |
| $\mathcal{B}^+/\mathcal{B}\,(\%)$ | 100 | 102 | 103 | 102 | 101 | 103 | 166 | 107 | 137 | 112 |
| $\text{nb}^+/\text{nb}_{\text{tot}}\,(\%)$ | 25 | 35 | 18 | 23 | 26 | 6.4 | 0.13 | 7.7 | 0.40 | 0.39 |
| $E^a(\%)$ | 5.4 | 5.0 | 4.7 | 4.3 | 5.4 | 4.6 | 0.074 | 2.7 | 2.7 | 4.5 |
| $L_T(\%)$ | 2.9 | 2.7 | 3.4 | 2.8 | 1.6 | 2.7 | 3.4 | 5.6 | 11 | 64 |

TABLE 1. Summary of simulation parameters (top part of the table) and time-averaged output metrics (in the quasi-steady regime) related to mixing (bottom part), defined throughout this paper. The reader is referred to table 1 in Pt. 1 for spatial and temporal discretizations. Powers $\mathcal{P}$ and $\mathcal{B}$ are normalized by the production of kinetic energy of the base flow $\mathcal{P}_{\text{lam}}$ (3.18). The available potential energy $E^a$ is normalized by the kinetic energy of the base flow.

The potential energy

$$E^p := \beta \iiint_{\mathcal{V}} \rho z \, \mathrm{d}\mathcal{V} \qquad (3.5)$$

budget differs from the usual formula for a closed system, as we shall see, because of axial periodicity and the presence of the linear forcing term (2.4) for stratification. There are three terms

$$\frac{\mathrm{d}E^p}{\mathrm{d}t} = \mathcal{B} - \mathcal{F}_{\text{adv}} - \mathcal{F}_{\text{diff}}, \qquad (3.6)$$

where

$$\mathcal{F}_{\text{adv}} := \beta \oiint_A \rho z \mathbf{u} \cdot \mathbf{n} \, \mathrm{d}A \qquad (3.7)$$

$$= \beta V \langle \rho w \rangle_{\text{top/bottom}}, \qquad (3.8)$$

$$\mathcal{F}_{\text{diff}} := -\frac{\beta}{Re\,Sc} \oiint_A z \boldsymbol{\nabla} \rho \cdot \mathbf{n} \, \mathrm{d}A \qquad (3.9)$$

$$= -\frac{\beta}{Re\,Sc} V \langle \partial_z \rho \rangle_{\text{top/bottom}} \qquad (3.10)$$

are fluxes related to advection and diffusion of mass through the top and bottom boundaries. The notation $\langle . \rangle_{\text{top/bottom}}$ refers to a spatial average over the cross-section $S$ of the annulus, evaluated at either end of the domain $z = \pm\Gamma/2$ (the result is the same because of axial periodicity).

The background potential energy is defined as

$$E^b := \beta \iiint_\mathcal{V} \rho z_* \, \mathrm{d}\mathcal{V}, \tag{3.11}$$

where $z_*(r, \theta, z, t)$ is the height associated with a fluid particle after adiabatic sorting of the density field into its stage of minimum gravitational potential energy. At any given time, this reference height can be obtained by integration of the probability distribution function (pdf) of the density field (Tseng & Ferziger 2001), using

$$\mathrm{d}z_R|_t = -\Gamma \, \mathrm{pdf}(\tilde\rho, t)\mathrm{d}\tilde\rho, \quad z_R(\max_\mathcal{V} \tilde\rho) = -\frac{\Gamma}{2} \quad \text{and} \quad z_* = z_R(\tilde\rho, t). \tag{3.12}$$

According to Winters *et al.* (1995), the background potential energy budget can be written as

$$\frac{\mathrm{d}E^b}{\mathrm{d}t} = \mathcal{M} - \mathcal{F}^b_{\mathrm{adv}} - \mathcal{F}^b_{\mathrm{diff}}, \tag{3.13}$$

where

$$\mathcal{M} := \frac{\beta}{Re\,Sc} \iiint_\mathcal{V} -\frac{\partial z_R}{\partial \tilde\rho} \|\boldsymbol{\nabla}\rho\|^2 \, \mathrm{d}\mathcal{V}, \tag{3.14}$$

$$\mathcal{F}^b_{\mathrm{adv}} := \beta \oiint_A \psi \mathbf{u} \cdot \mathbf{n} \, \mathrm{d}A, \quad \text{where} \quad \psi = \int^{\tilde\rho} z_R(\xi, t) \, \mathrm{d}\xi, \tag{3.15}$$

$$\mathcal{F}^b_{\mathrm{diff}} := -\frac{\beta}{Re\,Sc} \oiint_A z_* \boldsymbol{\nabla}\rho \cdot \mathbf{n} \, \mathrm{d}A. \tag{3.16}$$

The first term $\mathcal{M}$ is the irreversible dyapicnal mixing term while the second and third terms are advective and diffusive fluxes.

Using the potential and background potential energy budgets, one can derive an equation for the evolution of the available potential energy

$$E^a := E^p - E^b. \tag{3.17}$$

The significance of these different potential energies will be discussed further in §??.

In the following subsections, we will discuss properties of these energy budgets in a vertically periodic system with the example of run C.

### 3.2. *Stratification inhibits dissipation*

The kinetic energy budget in figure 2 features a transient evolution of all terms for $t \lesssim 400$, before saturating to quasi-steady values. All terms are normalized by the production of kinetic energy in the laminar base flow

$$\mathcal{P}_{\mathrm{lam}} = \frac{4\pi\Gamma}{Re(1-\eta^2)}, \tag{3.18}$$

therefore, $\mathcal{P}$ represents the ratio of the instantaneous torque to its laminar value, sometimes called Nusselt number $Nu_\omega$ (Eckhardt *et al.* 2007; Brauckmann & Eckhardt 2013). Our results are in line with values found in turbulent Taylor vortex flow in the unstratified case. Indeed, figure 11($a$) in Brauckmann & Eckhardt (2013) shows the evolution of $Nu_\omega$ against a shear Reynolds number,

$$Re_S := \frac{2}{1+\eta}Re, \tag{3.19}$$

for various values of the radius ratio $0.68 \leqslant \eta \leqslant 0.76$, including experimental and numerical studies from different research groups. This figure indicates that the Nusselt number increases from about 6 to 12 in the range of shear Reynolds numbers $3000 \leqslant Re_S \leqslant 14000$ spanned by our results. These values are perfectly consistent with ours, reported in table 1, which suggests that stratification only has a minor effect on the torque. This result is consistent with the radial profile of angular momentum being similar in the stratified case (figure 8(a) in Pt. 1) and in the unstratified case (figure 10(b) in Brauckmann & Eckhardt (2013)), given the close connection established by Brauckmann & Eckhardt (2013) between boundary layer thickness and torque. However, because of the positive buoyancy flux $\mathcal{B}$, there must be less dissipation in the stratified case than in the unstratified one, for the same input of kinetic energy. Therefore, turbulence is less active in the stratified case, as expected from the hindering of vertical motions by the restoring buoyancy force. The power that is not dissipated in the stratified case is used to carry mass upwards instead.

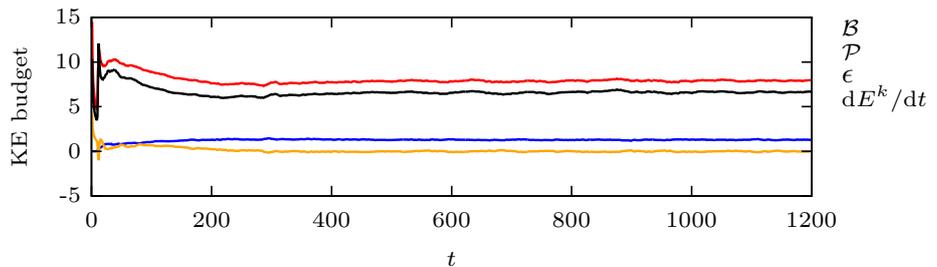

FIGURE 2. Times series of kinetic energy budget for run C (all terms are normalized by the for production of kinetic energy in the base flow (3.18)).

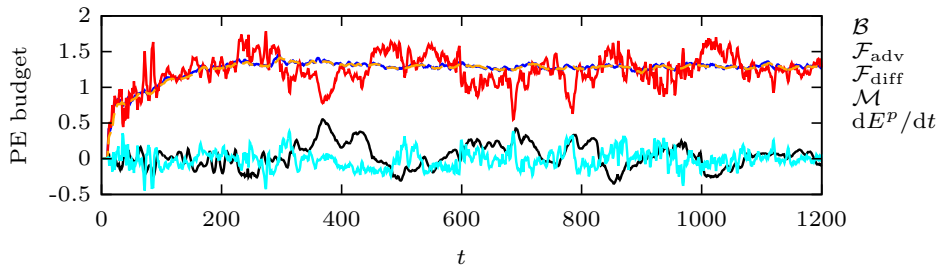

FIGURE 3. Times series of potential energy budget for run C, with the additionnal term $\mathcal{M}$ from the background potential energy budget (as in figure 2, all terms are normalized by the for production of kinetic energy in the base flow (3.18)).

### 3.3. *Intrinsic quantities related to potential energy: dyapicnal mixing, buoyancy flux and available potential energy*

Figure 3 shows every term in the potential energy budget, normalized again by $\mathcal{P}_{\text{lam}}$. Oscillations of $\mathcal{F}_{\text{adv}}$ around the value of $\mathcal{B}$ are an indication of stirring, which is a reversible process. However, the fact that $\mathcal{F}_{\text{adv}}$ averages to $\mathcal{B}$ rather than zero shows that surface fluxes do not exclusively represent reversible processes, as stated in Salehipour & Peltier (2015). As we shall see in §??, the buoyancy flux $\mathcal{B}$ is here a measure of an irreversible process, as it is nearly equal to the dyapicnal mixing term $\mathcal{M}$. The diffusive flux oscillates around zero here and therefore does not contribute to the potential energy budget on average. The slow oscillations of both flux terms are due to the vertical drift of the interfaces due to the broken symmetry of the mixed-ribbon ($\omega_{\text{axi}} \neq 0$, as explained in Pt. 1). For a pure ribbon, the temporal frequencies of the 'parent waves' are identical, and therefore there is no vertical drift of the axisymmetric modes. In that case, $\mathcal{F}_{\text{adv}}$ and $\mathcal{F}_{\text{diff}}$ would fluctuate very weakly around a fixed value which depends on the position of the axial boundaries with respect to the pattern. For instance, $\mathcal{F}_{\text{diff}} > 0$ and $\mathcal{F}_{\text{adv}} < \mathcal{B}$ if $z = \pm \Gamma/2$ is the position of an interface, and $\mathcal{F}_{\text{diff}} < 0$ and $\mathcal{F}_{\text{adv}} > \mathcal{B}$ if $z = \pm \Gamma/2$ is halfway between two horizontal interfaces.

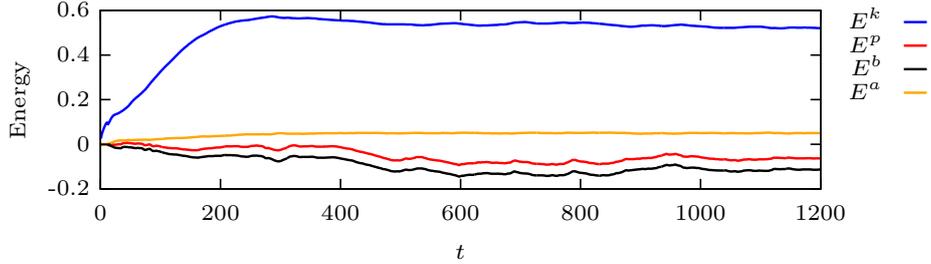

FIGURE 4. Times series of kinetic, potential, background potential and available potential energies for run C. All the energies are normalized by the kinetic energy of the laminar base flow.

The boundary fluxes are not the only terms in the potential energy budgets which depend on the position of the periodic pattern with respect to the axial boundaries. In fact, the potential and background potential energies themselves (see figure ??) would also change if the periodic pattern was translated vertically. Using the spectral decomposition of the density field, we can show that the potential energy is stored exclusively by the axisymmetric modes

$$E^p = \sum_{k>0} E^p_k, \quad \text{where} \quad E^p_k := 2\beta\Gamma^2 \frac{(-1)^k}{k} \int \text{Im}(\rho_{0,k}) r \, \mathrm{d}r. \quad (3.20)$$

The spectral coefficients $\rho_{0,k}$ would be replaced by $\rho_{0,k} \exp -\mathrm{i}k k_0 \delta z$ under a translation $z \to z + \delta z$ in the spectral expansion (2.12). Therefore, $E^p$ is sign-indefinite and the decrease in potential energy in figure 3 is physically meaningless: by choosing $\delta z$ appropriately, it is possible make $E^p > 0$. Note that this has nothing to do with translating the origin: we are translating the pattern, but not the origin of the domain $\mathcal{V}$ which is still centered about $z = 0$. Similarly, the counterintuitive decrease in background potential energy is also physically meaningless. As we shall see in §3.7, end-effects in a closed system would more than compensate any decrease in background potential energy from the interior, such that $\mathrm{d}E^b/\mathrm{d}t > 0$, as expected from Winters *et al.* (1995)'s theory.

The quantities which are intrinsic to the periodic system are those which do not change under an arbitrary translation of the pattern. For that reason, all the quantities involved in the kinetic energy budget are intrinsic. For instance, the fact that the kinetic energy of the flow is lower than that of the laminar Couette solution is physically meaningful. It is related to the strong mean flow distortion, leading to a nearly constant angular momentum profile across the annulus (see figure 8(a) in Pt. 1), except in boundary layers. The coherent vortical motions and turbulence lead to velocity rms which are lower than 6% of the inner cylinder velocity at any radial position (figures 8(b, d) in Pt. 1), therefore, these do not compensate for the loss of kinetic energy through the mean azimuthal velocity component. The meaningful quantities related to potential energy are the dyapicnal mixing term $\mathcal{M}$, the buoyancy flux $\mathcal{B}$ and the available potential energy $E^a$. To understand it, define the inner product

$$(f, g) := \iiint_{\mathcal{V}} fg^* \, \mathrm{d}\mathcal{V} \quad (3.21)$$

over the space of complex periodic functions of $\theta$ and $z$. Using the spectral decomposition (2.12) of $f$ and $g$, we have

$$(f, g) = \sum_{m,k} (f_{m,k}, g^*_{m,k}) \quad (3.22)$$

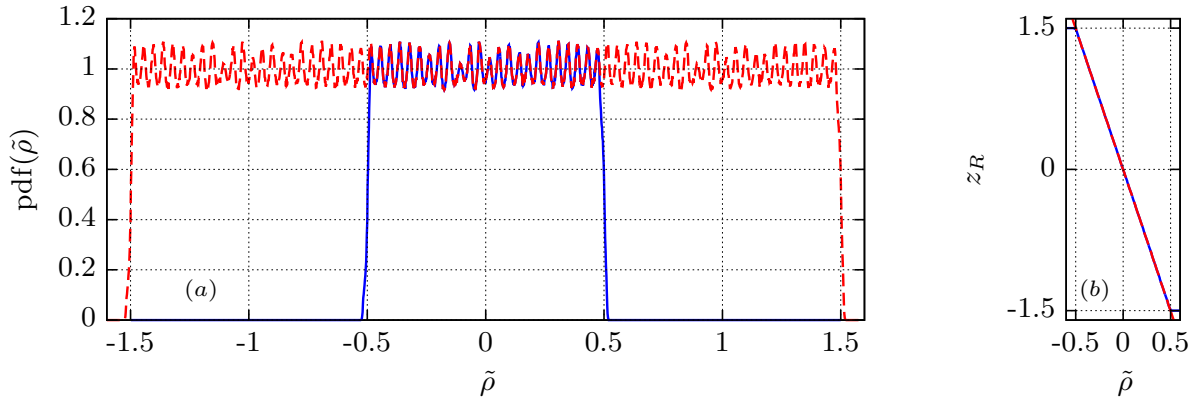

FIGURE 5. (*a*) Probability distribution function of density field of a snapshot at the end of simulation C. The solid line was obtained by sorting the density field within the computational box $-\Gamma/2 \leqslant z < \Gamma/2$ only. The dashed line was obtained by extrapolating the density field to $-3\Gamma/2 < z < 3\Gamma/2$, using periodicity (and renormalizing the resulting pdf by a factor of 3 to coincide with the solid line on most of the interval $-0.5 < \tilde{\rho} < 0.5$). (*b*) Background density profile $z_R$ computed from the two pdfs, using (3.12) (same nomenclature as (*a*) for solid/dashed lines).

(Parseval's identity), with the additionnal property that $(f_{m,k}, g^*_{m,k}) = (f_{-m,-k}, g^*_{-m,-k})$ if $f$ and $g$ are real fields. The aforementioned quantities can all be written under the form of this inner product

$$\mathcal{B} = \beta(\rho, w), \qquad (3.23)$$

$$\mathcal{M} = \frac{\beta \Gamma}{Re\, Sc}(\text{pdf} \circ \tilde{\rho}, \|\boldsymbol{\nabla}\rho\|^2), \qquad (3.24)$$

$$E^a = \beta(\rho, \zeta), \qquad (3.25)$$

where $\zeta := z - z_*$ is a $\theta, z$-periodic vertical displacement field and $\circ$ denotes composition. As a result, all these quantities are invariant under an arbitrary translation along $z$: the phase shift induced on $f_{m,k} \to f_{m,k} \exp{-ikk_0\delta z}$ is exactly compensated by an opposite phase shift on $g^*_{m,k} \to g^*_{m,k} \exp{+ikk_0\delta z}$.

In fact, the last two identities are only approximately true if the density field is sorted over the finite volume $\mathcal{V}$. Indeed, the fields pdf $\circ\ \tilde{\rho}$ and $\zeta$ are truly periodic only in the limit of an infinite number of axial periods (pdf$[\tilde{\rho}(r,\theta,z+\Gamma)] = $ pdf$[\tilde{\rho}(r,\theta,z) - 1] = $ pdf$[\tilde{\rho}(r,\theta,z)]$ because pdf$[\tilde{\rho}]$ is 1-periodic over an infinite domain). However, as can be seen in figure 5, the pdf of the density field sorted over 3 periods is an excellent approximation of the pdf over one period only, when pdf $\approx 1$ (pdf $\sim 1/3$ when $\Gamma \to 3\Gamma$ because of the definition of the reference density scale $\Delta\rho \to 3\Delta\rho$). A small error is introduced by the 'tails' of the pdf, when $\tilde{\rho}$ is close to either its maximum or minimum value. But we can always neglect this error by taking an appropriately large number of axial periods to define $\mathcal{V}$, if pdf $\not\approx 1$: then the three metrics $\mathcal{B}$, $\mathcal{M}$ and $E^a$ become truly intrinsic to mixing.

This is why available potential energy has to be positive, even in an open system, as it is necessary for stirring hence mixing to occur. Any lateral variations in $\tilde{\rho}$ or statically unstable regions (figure 4 in Pt. 1) guarantees that $E^a > 0$ (Peltier & Caulfield 2003). However, the values of $E^a$ (computing the pdf of $\tilde{\rho}$ over 3 axial periods instead of 1) in table 1 are always very small compared to the kinetic energy of the base flow. We will see the implication in the next section.

### 3.4. *Mixing equals buoyancy flux in the limit of small Thorpe scale*

The vertical displacement field $\zeta$ can be characterised by its rms value, or Thorpe scale

$$L_T := \sqrt{\frac{1}{\mathcal{V}} \iiint_\mathcal{V} \zeta^2 \, \mathrm{d}\mathcal{V}}. \tag{3.26}$$

This length scale (computing the pdf of $\tilde{\rho}$ over 3 axial periods instead of 1) is tabulated in table 1, together with available potential energy $E^a$. In all cases A–G, $L_T$ is only a few percent of the gap width and $E^a$ is only a few percent of the kinetic energy of the base flow. In that case, we can assume that $z_* \approx z$, and therefore

$$E^p \approx E^b, \quad F_{\mathrm{adv}} \approx F^b_{\mathrm{adv}}, \quad F_{\mathrm{diff}} \approx F^b_{\mathrm{diff}} \tag{3.27}$$

leading to

$$\mathcal{B} \approx \mathcal{M}. \tag{3.28}$$

In other terms, in the limit of vanishing Thorpe scale/available potential energy, we can identify potential and background potential energies, hence mixing equals buoyancy flux. This is what we observe in figure 4, at all times.

Furthermore, consider the evolution equation

$$\frac{\mathrm{d}\sigma}{\mathrm{d}t} = \frac{2}{\beta \Gamma \mathcal{V}}(\mathcal{B} - \epsilon_P). \tag{3.29}$$

for the perturbation density variance

$$\sigma := \frac{1}{\mathcal{V}} \iiint_\mathcal{V} \rho^2 \, \mathrm{d}\mathcal{V}, \tag{3.30}$$

where

$$\epsilon_P := \frac{\beta \Gamma}{Re\, Sc} \iiint_\mathcal{V} \|\boldsymbol{\nabla}\rho\|^2 \, \mathrm{d}\mathcal{V} \tag{3.31}$$

(a quantity sometimes called dissipation of potential energy (ref??)). In the quasi-steady regime, we obviously have $\mathcal{B} \approx \epsilon_P$. So if additionnally $L_T \ll 1$, we then have $\mathcal{M} \approx \epsilon_P$, leading to the following integral condition on the pdf of $\tilde{\rho}$

$$\iiint_\mathcal{V} [\mathrm{pdf}(\tilde{\rho}) - 1] \|\boldsymbol{\nabla}\rho\|^2 \, \mathrm{d}\mathcal{V} \approx 0. \tag{3.32}$$

Having a pdf nearly constant and equal to one, and hence an almost linear background stratification (as seen in figure 5) is consistent with this condition, although it is not necessary for (3.32) to be true. The assumption that $z_* \approx z$ does not imply that the background stratification should be linear, it only means that the flow is close to its state of minimum potential energy.

### 3.5. *The more disordered the flow, the lower mixing efficiency*

Given the fact that the buoyancy flux is here a direct measure of mixing (cf. equation (3.28)), an appropriate measure of mixing efficiency in this problem is the flux Richardson number defined as

$$Ri_f := \frac{\mathcal{B}}{\mathcal{P}}. \tag{3.33}$$

Figure 6 shows time series of $Ri_f$ for the different simulations, and average values are reported in table 1, together with those of $\mathcal{P}$ and $\mathcal{B}$. We observe significant scatter of the mean values of $Ri_f$, which increases with $Ri$ but decreases with all the other parameters:

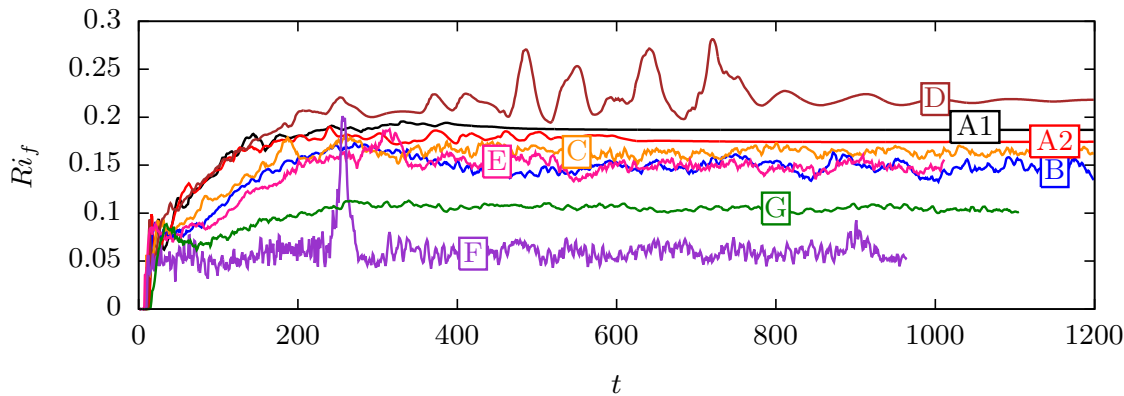

FIGURE 6. Time series of $Ri_f$ for all simulations.

$\eta$, $Sc$, $Re$. The effect of $Re$ is weak though, so long as we consider flows dominated by the same type of coherent structure (indeed, ribbon states lead to larger $Ri_f$ than mixed-ribbons: A1 versus A2 for the same parameters, or the particularly large value of $Ri_f$ for run D). The dependence of $Ri_f$ with the different control parameters seem to obey a simple rule: the more disordered the flow is, the less efficiently it mixes. Indeed, with greater dissipation, there is less energy available for stirring hence mixing. The time series give a perception of the range of time-scales involved in each simulation and confirm the negative impact of turbulence on $Ri_f$. Large fluctuations in $Ri_f$ for simulations D (large $Ri = 10$) and F ($Sc = 10$) correspond to transient events that can be interpreted as visits of different coherent states (see also spatiotemporal diagram for run D in figure 7 of Pt. 1). Whether the sudden surges in $Ri_f$ are linked to the properties of the visited saddle, or to the transient nature of the dynamics is unclear. However, since mixed-ribbons have lower $Ri_f$ than ribbons, and since ribbons and purely helical branches seem to have similar $Ri_f$ values, as we will see in §3.6, the large fluctuations in $Ri_f$ seem more likely to be due to transient effects.

### 3.6. *Mass transport by nonlinear waves*

To investigate the origin of the buoyancy flux, we consider its spectral decomposition, already introduced in §3.3

$$\mathcal{B} = \sum_{m,k} \mathcal{B}_{m,k}, \quad \mathcal{B}_{m,k}(r,t) := \beta(\rho_{m,k}, w_{m,k}) \tag{3.34}$$

$$= 2\pi \Gamma \beta \int |\rho_{m,k}||w_{m,k}| \cos \Delta\Phi_{m,k} r \, \mathrm{d}r, \tag{3.35}$$

where $\Delta\Phi_{m,k}(r,t)$ denotes the phase shift between the density and vertical velocity components of mode $(m,k)$, at radial position $r$ and time $t$. Each term $\mathcal{B}_{m,k}$ can be seen as the *active power* consumed by mode $(m,k)$ for the transport of mass. We can therefore define an associated *instantaneous power factor*

$$\lambda_{mk} := \frac{(\rho_{m,k}, w_{m,k})}{\|\rho_{m,k}\|\|w_{m,k}\|}, \tag{3.36}$$

by analogy with electric dipoles in alternating current. At a given radial position and time, a mode contributes positively to the buoyancy flux (upwards mass flux) if $|\Delta\Phi_{m,k}| < \pi/2$, e.g. if $\rho_{m,k}$ and $w_{m,k}$ are positively correlated. Equivalently, a spatial mode is globally contributing to the buoyancy flux at a given time $t$ if its power factor is positive. We stress the fact that the convenient decomposition (3.35) is energetic in nature and tells nothing about the mechanisms of Lagrangian transport, which also involve phase shifts between the different modes. Finally, we note that $\mathcal{B}_{-m,-k} = \mathcal{B}_{m,k}$ because of Hermitian symmetry. The definition of the power factor allows a scale-by-scale analysis of the correlation coefficient between density and vertical velocity fields introduced by Ivey *et al.* (1998). We can also dissociate the portion of the spectrum contributing positively to $\mathcal{B}$, from that contributing negatively. The fraction of modes transporting mass up (resp. down) will be denoted $\text{nb}^+/\text{nb}_\text{tot}$ (resp. $\text{nb}^-/\text{nb}_\text{tot}$), and the corresponding buoyancy flux $\mathcal{B}^+$ (resp. $\mathcal{B}^-$).

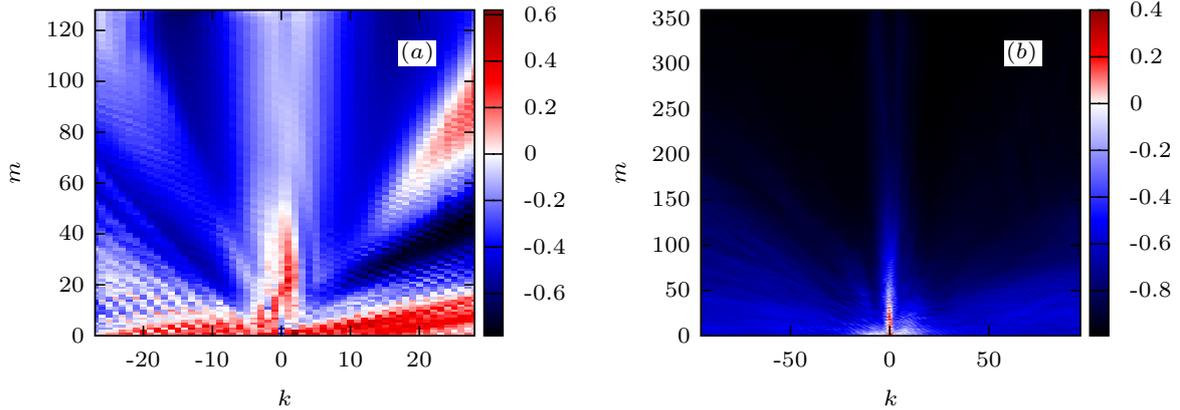

FIGURE 7. Time-averaged power factor spectra for (a) run C ($Sc = 1$) and (b) run F ($Sc = 10$), both with $\Gamma = 3/8$.

Figure 7 shows the spectrum of power factor averaged over time, for runs C and F, corresponding to two different values of $Sc$, respectively 1 and 10 (we checked that the metrics in table 1 do not change when the size of box is reduced from $\Gamma = 3$ to $\Gamma = 3/8$ for run C, except for the dominant modes which become $(2, 1)$ and $(1, -1)$ indeed. For simulation C at $\Gamma = 3/8$, the resolution was $84 \times 256 \times 56$.). In both cases, it appears that modes contributing positively, on average, to the buoyancy flux, correspond to large scale coherent structures. Small scales, on the other hand, contribute negatively to $\mathcal{B}$ for the most part. This is even more striking for $Sc = 10$ than for $Sc = 1$, although finer turbulent structures exist. For $Sc = 10$ indeed, only $\text{nb}^+/\text{nb}_\text{tot} = 0.13\%$ of the modes contribute positively to $\mathcal{B}$. However, these modes are so energetic that they are nonetheless able to lead to an upwards flux. Since the rest of the modes are transporting $\mathcal{B}^-/\mathcal{B} = 2/3$ in the 'wrong' direction, this requires the nonlinear waves to contribute $\mathcal{B}^+/\mathcal{B} \approx 5/3$ upwards. Therefore, the mechanism of 'scouring' of interfaces by turbulent eddies of Oglethorpe *et al.* (2013) does not seem to apply to our low-$Sc$ simulations: if anything, turbulence is in fact moving mass down, and this effect is getting worse as the Schmidt number increases from 1 to 10. Similarly, increasing the value of the Reynolds number from $Re = 2000$ to 5000, then 10000, has a negative impact on the power factors, leading to a decrease in the proportion of modes transporting mass up: respectively $\text{nb}^+/\text{nb}_\text{tot} = 35\%, 18\%, 6.4\%$ for simulations A2, C and E. But the flux Richardson number is not dramatically impacted, although slightly decreasing. Again, the flow in the smaller gap $\eta = 0.625$ is more disordered than when $\eta = 0.417$, and $\text{nb}^+/\text{nb}_\text{tot}$ drops from 23% (run C) to only 7.7% (run G).

Figure 8(a,b) shows the spectrum of buoyancy flux for modes contributing more than 1% of $\mathcal{B}$ in absolute value in runs D and C. All the selected modes contribute positively since there is no single negative contribution amounting more than 1% of $\mathcal{B}$. In both cases, only 4 modes (+ complex conjugates) account for approximately 90% of $\mathcal{B}$ (and roughly 90% of $\mathcal{B}^+$ as well since $\mathcal{B}^+/\mathcal{B} \gtrsim 100\%$ for C and D), and those correspond to the

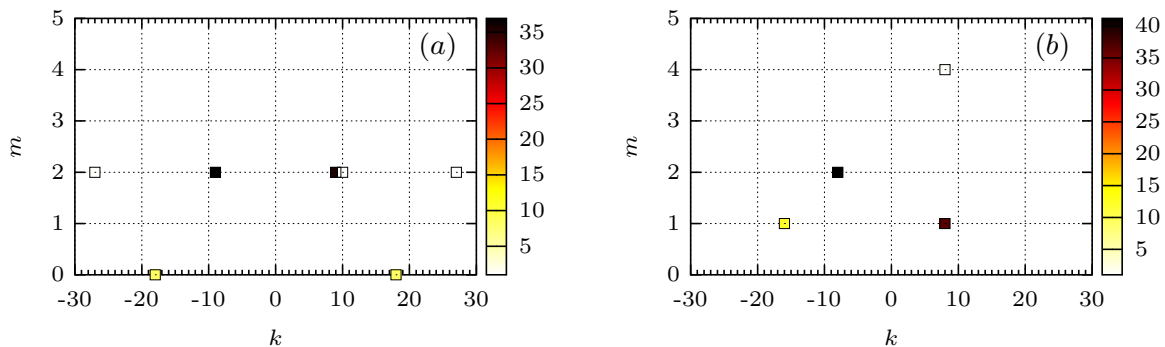

FIGURE 8. Time-averaged buoyancy flux spectrum, normalized by the total buoyancy flux $\mathcal{B}$, in percentage. To take into account complex conjugate modes, which have the same buoyancy flux, we plot $\mathcal{B}_{m,k} + \mathcal{B}_{-m,-k} = 2\mathcal{B}_{m,k}$ when $m \neq 0$, and just $\mathcal{B}_{0,k}$ otherwise. Only modes contributing more than 1% of the total buoyancy flux, in absolute value, are plotted (no mode appears to give a significant negative contribution to $\mathcal{B}$). (a) Run D, (b) run C. In both cases, the handful of modes account for about 90% of $\mathcal{B}$.

most energetic helical modes in the flow (see figure 3 in Pt. 1 showing the spectrum of kinetic energy for the same simulations). Indeed, among the few modes which carry mass upwards, only a very limited number provide a significant contribution to $\mathcal{B}$ because of the rapid decrease in amplitude of the spectral coefficients with $m$ and $k$ in equation (3.35).

To further understand the coupling between the density and vertical velocity components, we examine the equation of advection-diffusion of a finite-amplitude spatial mode $\mathbf{q}_{m,k}(r,t)$. To this end, we separate the velocity field $\mathbf{u}$ into the basic Couette flow $V(r)\mathbf{e}_\theta$ and a perturbation $\mathbf{u}' := \mathbf{u} - V(r)\mathbf{e}_\theta$. After a Laplace transform in time $\hat{\mathbf{q}}_{m,k}(r,\omega) := \int_0^\infty \mathbf{q}_{m,k}(r,t)\exp[\mathrm{i}\omega t]\mathrm{d}t$, the equation becomes

$$\mathrm{i}(m\frac{V}{r} - \omega)\hat{\rho}_{m,k} - \frac{\hat{w}_{m,k}}{\Gamma} = -(\widehat{\mathbf{u}' \cdot \boldsymbol{\nabla}\rho})_{m,k} + \frac{1}{Re\,Sc}\left(\partial^2_{rr} + \frac{1}{r}\partial_r - \frac{m^2}{r^2} - (kk_0)^2\right)\hat{\rho}_{m,k} \tag{3.37}$$

(note that at $t = 0$, $\rho_{m,k} = 0$ as we only perturb the velocity field initially). If mode $\mathbf{q}_{m,k}$ is simply a wave of real frequency $\omega$, then $\mathbf{q}_{m,k} = \hat{\mathbf{q}}_{m,k}\exp[-\mathrm{i}\omega t]$ and the phase shift between $\rho_{m,k}$ and $w_{m,k}$ is at all times equal to the phase shift between $\hat{\rho}_{m,k}$ and $\hat{w}_{m,k}$. Because $\omega$ is real, the coupling (i.e. $|\Delta\Phi_{m,k}| \neq \pi/2$) is possible only if the right-hand-side of (3.37) is non-zero, there are therefore two possible sources: advection by the perturbation velocity $\mathbf{u}'$ and diffusion. To distinguish between these effects, we use continuation to follow the ribbon and helical branches bifurcating out of the base flow as the Reynolds number increases.

Figure 9(a) shows the evolution of the power factor associated with the dominant helical mode(s), for three different values of the Schmidt number: $Sc = 1, 10, 100$ and $Ri = 3$. At criticality (leftmost point on each curve), the nonlinear term in (3.37) is zero so the coupling is only due to diffusion. It is therefore not surprising to find that the power factor decreases with $Sc$ then. However, as $Re$ increases, we find different trends for $\lambda_c(Re)$: the power factor decreases for $Sc = 1$, but is able to increase for $Sc = 10$ and $Sc = 100$ (even though the evolution is non-monotonic for the purely helical branch at $Sc = 10$). So diffusive and advective effects combine in a non-trivial way to induce a strictly positive coupling beyond the instability threshold.

Another remarkable feature of the bifurcation analysis is the rapid saturation of $Ri_f$ in figure 9(b) to a value close to that found in the DNS, at much larger $Re$, and with many more modes in the flow. Indeed, $Ri_f$ approaches 0.16 for $Sc = 1$ and 0.06 for $Sc = 10$, which is consistent with the corresponding values in figure 6 and table 1 for a range of Reynolds numbers. We checked that this was not mere coincidence and it worked equally well for run G at $\eta = 0.625$. This proves, again, that the dynamics of mixing is dominated by the large coherent structures of the flow. Also, the prediction of $Ri_f$ by analysis of the coherent motions seems robust, in the sense that it does not strongly depends on the details of structure selected by the flow: helical and ribbon branches yield similar $Ri_f$ in figure 9(b) and this is also true for ribbons and mixed-ribbons in simulations A1 and A2 (figure 6). Moreover, the nonlinear branches created from the critical mode in figure 9 do not correspond to the actual coherent structures selected in the DNS, which have different wavenumbers, yet they yield excellent estimates for $Ri_f$.

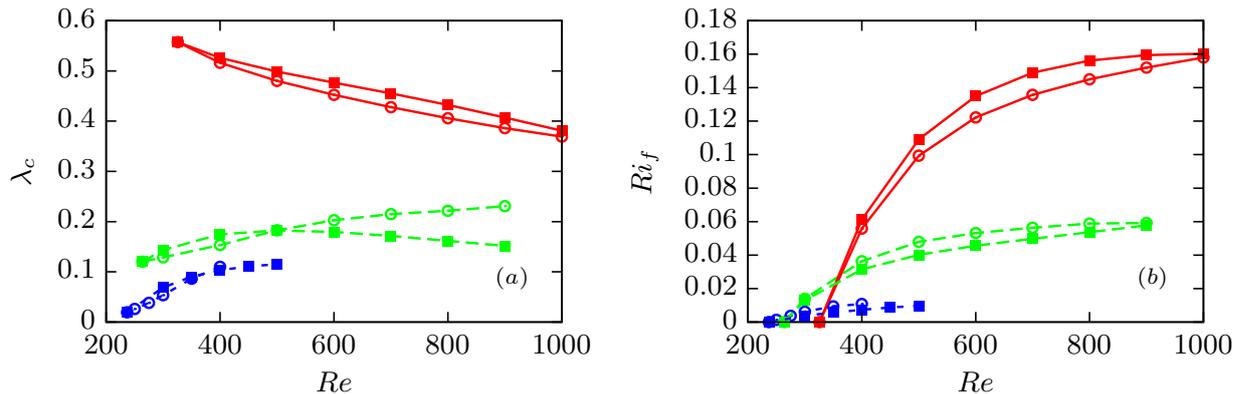

FIGURE 9. (a) Bifurcation diagram of the power factor $\lambda_c$ versus $Re$ associated with the dominant helical mode(s) of ribbon (LH/RH interaction; open symbols) and helical (LH+harmonics or RH+harmonics; solid symbols) branches created by the critical instability mode at $Ri = 3$ and $Sc = 1, 10, 100$ (resp. solid, dashed and dotted lines). (b) Corresponding bifurcation diagrams for $Ri_f$.

Finally, in figure 10, we compare the radial profiles of local buoyancy flux $b := g\rho w$ (averaged over $\theta$, $z$ and time) associated with (a) the linear critical mode and (b) the total flow at $Re = 5000$. There are striking similarities between the two figures. In both cases, the buoyancy flux is always/nearly positive everywhere for the linear/nonlinear flow, which indicates a positive correlation between $\rho$ and $w$ at all/nearly all radial positions. The local buoyancy flux is peaked near the boundaries, with a smaller contribution near the outer cylinder for the larger gap case: mass is transported upwards preferentially near the boundaries, where the waves are localized (see the distribution of $|w|$ for the critical linear mode in figure 2 of Pt. 1). The most notable difference between (a) and (b) is the presence of sharper boundary layers at larger $Re$ in the nonlinear flow. In particular, there is a small region of negative buoyancy flux in the vicinity of the inner cylinder which does not seem to be captured by linear theory (although $Re$ is only $O(10^2)$ in (a)). Apart from these small differences, the local buoyancy flux distribution is well-captured by the critical linear mode, which is another indication that the mechanism for mass transport is essentially linear at $Sc = 1$.

### 3.7. *The role of end-plates in the energetics of mixing*

In this section, we come back to the analysis of §3.1 and discuss the effect of the end-plates in the potential energy budget. In Winters *et al.* (1995), there is no assumption on the form of the perturbation density field $\tilde{\rho}$, and the potential energy budget is written as

$$\frac{\mathrm{d}\widetilde{E^p}}{\mathrm{d}t} = \widetilde{\mathcal{B}} - \widetilde{\mathcal{F}}_{\mathrm{diff}} - \widetilde{\mathcal{F}}_{\mathrm{adv}} + \widetilde{\mathcal{D}_p}, \quad (3.38)$$

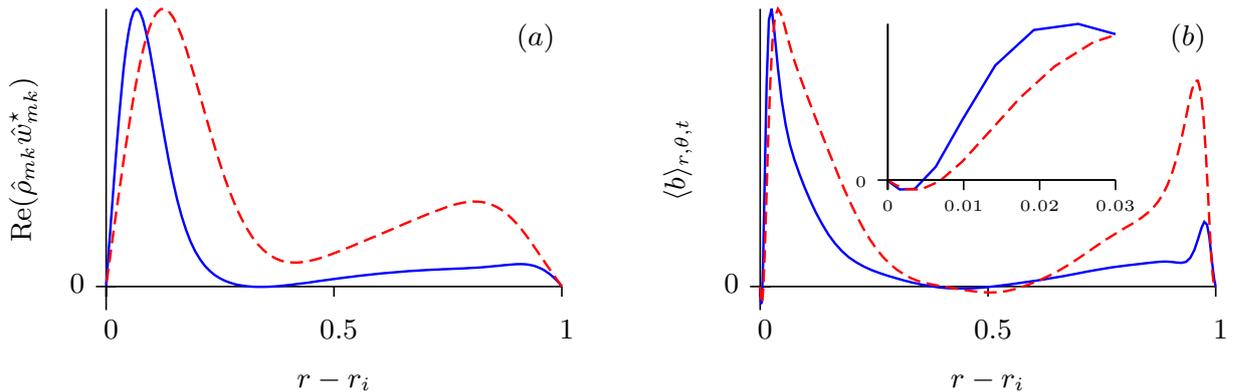

FIGURE 10. (a) Integrand of the modal buoyancy flux (3.35) for the critical mode at $Ri = 3$, $Sc = 1$. Solid line: $\eta = 0.417$ ($m = 1$). Dashed line: $\eta = 0.625$ ($m = 2$). (b) Radial profile of local buoyancy flux $b := \beta \rho w$ (normalized by its maximum value), averaged over the two periodic directions and time, for simulations C and G at respectively $\eta = 0.417$ and $\eta = 0.625$ (same line styles as (a)), $Ri = 3$, $Sc = 1$ and $Re = 5000$.

where
$$\widetilde{E^p} = \beta \iiint_\mathcal{V} \tilde{\rho} z \, \mathrm{d}\mathcal{V}, \quad \widetilde{\mathcal{B}} = \beta \iiint_\mathcal{V} \tilde{\rho} w \, \mathrm{d}\mathcal{V} \quad (3.39)$$
are the potential energy and buoyancy flux definitions based on $\tilde{\rho}$,
$$\widetilde{\mathcal{F}}_\mathrm{adv} = \beta \oiint_A \tilde{\rho} z \mathbf{u} \cdot \mathbf{n} \, \mathrm{d}A, \quad \widetilde{\mathcal{F}}_\mathrm{diff} = -\frac{\beta}{Re\,Sc} \oiint_A z \boldsymbol{\nabla} \tilde{\rho} \cdot \mathbf{n} \, \mathrm{d}A. \quad (3.40)$$
are advective/diffusive fluxes also based on $\tilde{\rho}$ and
$$\widetilde{\mathcal{D}_p} = -\frac{\beta S}{Re\,Sc}(\langle \tilde{\rho} \rangle_\mathrm{top} - \langle \tilde{\rho} \rangle_\mathrm{bottom}) \quad (3.41)$$
is a term representing the rate of conversion from internal to potential energy.

It is remarkable that the term $\widetilde{\mathcal{D}_p}$ disappears from the potential energy budget (3.6) when the perturbation density field $\tilde{\rho}$ is decomposed into a linear part $\bar{\rho}$ and a periodic perturbation $\rho$. To further understand this, decompose every term in (3.38) as a sum of two terms $\widetilde{(.)} = \overline{(.)} + (.)$, the first one depending on $\bar{\rho}$ while the second depends on some perturbation $\rho$, not necessarily periodic. It follows immediately that
$$\frac{\mathrm{d}\overline{E^p}}{\mathrm{d}t} = \overline{\mathcal{B}} = \overline{\mathcal{F}_\mathrm{adv}} = 0, \quad (3.42)$$
and
$$\overline{\mathcal{F}_\mathrm{diff}} = \overline{\mathcal{D}_p}, \quad (3.43)$$
meaning that the largest contribution $\overline{\mathcal{D}_p}$ to $\widetilde{\mathcal{D}_p}$ is cancelled by a boundary term. This occurs because the density difference between the top and bottom ends of the domain is maintained at all times. The remaining part $\mathcal{D}_p$ is generally small and exactly zero if $\rho$ is assumed to be periodic, leading to equation (3.6) for the time-evolution of the potential energy.

On the other hand, if we consider a closed volume $\mathcal{V}$ with no-flux boundary conditions, the density difference will evolve over time under the effect of diffusion, leading to an increase in potential energy quantified by $\widetilde{\mathcal{D}_p}$. At $t = 0$, this term would be equal to
$$\overline{\mathcal{D}_p} = \frac{Ri}{4Sc}\frac{(1-\eta^2)^2}{\eta^2}, \quad (3.44)$$

when normalized by the kinetic energy production in the base flow. For case C, this expression gives a value of $\overline{\widetilde{\mathcal{D}_p}} \approx 2.9$ which is larger than the buoyancy flux $\widetilde{\mathcal{B}} \approx 1.3$ (in the same unit). The value of $\overline{\widetilde{\mathcal{D}_p}}$ is relatively large because of the small Schmidt number, leading to rapid density diffusion. Yet, the mechanism of mixing in the interior of the flow is not expected to be different in the presence of end-plates. Indeed, despite its seemingly non-local expression, the term $\widetilde{\mathcal{D}_p}$ characterises diffusive processes which are initially confined to the vicinity of the end-plates, i.e. the thickening of homogeneous density layers created by the no-flux boundary condition. Whether small or large, this term does not quantify mixing far from the end-plates, and does not depend on the flow regime, i.e. laminar or turbulent.

Top and bottom plates also play another specific role: they stop the advective mass flux because of the impermeability condition, leading to accumulation/depletion of mass at the top/bottom of the apparatus. This also leads to the growth of top and bottom homogeneous layers near the end-plates, visible in the spatiotemporal conductivity diagrams of Oglethorpe *et al.* (2013), and results in an additional increase in potential energy. Indeed, whereas $\widetilde{\mathcal{B}}$ is cancelled by $\widetilde{\mathcal{F}}_{\text{adv}}$ on average in figure ??, this would not the case in a closed volume where $\widetilde{\mathcal{F}}_{\text{adv}} = 0$.

To summarize, realistic boundary conditions lead to transient dynamics until the flow becomes fully unstratified, which induces an (irreversible) increase in potential energy. This translates into the emergence of the term $\widetilde{\mathcal{D}_p}$ and the absence of an advective flux term in the potential energy budget of a closed system. Similar changes occur in the background potential energy budget, ensuring that the background potential energy always increase in a closed volume (Winters *et al.* 1995).

We finally add that in the case of a pure laminar ribbon (simulation A1), the pdf of the density field within the control volume does not change with time, but buoyancy flux hence mixing does occur. This is an indication that changes in the pdf are not required for mixing to occur, when considering an open control volume. This situation is also in contrast with closed systems, where mixing necessarily translates into an evolving pdf (Winters *et al.* 1995).

## 4. Discussion

In this section, we first use the buoyancy Reynolds number in §4.1 to discuss the relevance of turbulence in the mixing mechanism identified in this paper. Then, we revisit Oglethorpe *et al.* (2013)'s buoyancy flux measurements in the light of these findings, and try to evaluate whether or not our transport mechanism could be compatible with the observations.

### 4.1. *Turbulent versus chaotic mixing: buoyancy Reynolds number*

The buoyancy Reynolds number

$$Re_b := \frac{\epsilon}{\nu N^2}, \tag{4.1}$$

is a natural quantity to consider in the context of mixing in stratified turbulence. This number characterises the separation between the Kolmogorov scale $l_K$ and the smallest scale influenced by buoyancy, or Ozmidov scale $l_O$, such that $Re_b = (l_O/l_K)^{4/3}$. It is classically considered that flows with a buoyancy Reynolds number smaller than 20 do not have a large enough separation of scales to be in the strongly stratified turbulence regime Smyth & Moum (2000). It is also usually found that flows with either too large or too small $Re_b$ have asymptotically small mixing efficiency, as recently stated in a recent review article by Ivey *et al.* (2008): 'both laboratory and DNS work indicate that [...] when either $Re_b \sim O(1)$ or $Re_b \sim O(10^5)$, the mixing efficiency $Ri_f \to 0$ and the use of large $Ri_f \approx 0.2$ in field situations in these limits cannot be justified. This is not simply a matter of curiosity. There is a fundamental inconsistency between the results from the laboratory and DNS experiments and the inference of diffusivity from microstructure in the field that remains unresolved.' It is therefore remarkable that simulations A–G reach such large values of $Ri_f$, as high as 0.22 on average in case D, despite having $Re_b$ always of the order of 1, except for the smaller-gap case $\eta = 0.625$ where $Re_b = 12$ (see table 1). These high values of $Ri_f$ at low $Re_b$ seem possible precisely because the mass transport mechanism does not rely on turbulence, but waves. The efficient coupling between density and vertical velocity components at low $Sc$ is a linear mechanism which is key in justifying the large mixing efficiency.

In fact, the excellent mixing properties of coherent structures like ribbons should not come as a surprise if we refer to the Lagrangian analysis of Ashwin & King (1997): a study of particle paths in non-axisymmetric Taylor–Couette flows. Those authors studied the 'skeleton' of the flow in the corotating frame of the ribbon, identifying 'stagnation points and periodic orbits whose stable and unstable manifolds intersect in a complicated manner in the flow interior', hence creating large regions of chaotic mixing. This mechanism of chaotic mixing is entirely responsible for the upwards mass flux in our system, since turbulence carries buoyancy in the 'wrong' direction. We note a potential connection with another 'waveguiding flow', the stratified plane Couette flow, quoting Deusebio *et al.* (2015): 'remarkably, we find that although fluctuations are greatly suppressed in the laminar regions, $Ri_f$ does not change significantly compared with its value in turbulent regions'.

In simulation CT at $Sc = 16$, we were able to reach a buoyancy Reynolds number of $Re_b \approx 124 \gg 1$, but still only 0.39% of the spatial modes induce a positive buoyancy flux in this case, and these clearly correspond to low wavenumbers, i.e. coherent motions. The spectrum of power factor is not shown here, since it looks extremely similar to figure 7(*b*): only an extremely limited number of powerful modes are able to provoke upwelling, against the rest of the spatial scales. Again, we must conclude that upwards buoyancy flux is not caused by turbulence but by coherent motions, and that $Re_b$ is irrelevant to this process. The value of $Ri_f$ seems in fact much more sensitive to the Schmidt number than to $Re_b$, as can be seen when comparing C and F for instance: an increase by a factor of 10 in $Sc$ lead to a decrease by almost a factor of 3 in $Ri_f$. Similarly, simulations S and CT at $Sc = 16$ have comparable values of $Ri_f$ despite a very large difference in $Re_b$ between them (the first simulation with $Re_b = 5.4$ is laminar whereas the second with $Re_b = 124$ is turbulent), and both values of $Ri_f$ are small compared to the $Sc = 1$ simulations.

It may seem surprising at first to obtain such low values of $Re_b = O(1)$ even at $Re = 10000$ (run F) for all $\eta = 0.417$ cases, whereas CT at $\eta = 0.769$ reached a value of $Re_b > 100$ with only $Re < 1000$. This can be understood by writing $Re_b$ as

$$Re_b = \frac{4\epsilon\eta^2}{Ri(1-\eta^2)^2} \tag{4.2}$$

(where $\epsilon$ is normalized by $\mathcal{P}_{\text{lam}}$). Because $Ri_f \ll 1$, the dissipation $\epsilon$ is almost equal to $\mathcal{P}$ in the quasi-steady regime. And because stratification does not seem to have a great impact on $\mathcal{P}$, we can estimate this term to be lower than 10 for $Re < 10000$ (Brauckmann & Eckhardt 2013), therefore $\epsilon < 10$. This shows the limited impact $Re$ has on $Re_b$, compared to that of $\eta$ and $Ri$. The only noticeable correlation between $Re_b$ and mixing seems to lie in the density standard deviation (cf. Pt. 1), otherwise, buoyancy flux metrics $Ri_f$, $\mathcal{B}^+/\mathcal{B}$ and $\text{nb}^+/\text{nb}_{\text{tot}}$ seem to depend primarily on the Schmidt number.

### 4.2. *Universal flux law of Oglethorpe* et al. *(2013)*

In addition to initially linear stratification, Oglethorpe *et al.* (2013) also considered the case where there are five layers in the density profile at $t = 0$. Using conductivity measurements, they estimated the buoyancy flux in their experiments and reported on a universal flux law, governing the evolution of $\mathcal{B}$ against the Richardson number associated with a density interface. The buoyancy flux curve features a maximum, before decreases to an asymptotic value at larger $Ri$, which appears to be the same regardless of the form of the initial stratification. Our aim in this section is not to reproduce this law numerically, but merely to check whether the values of the buoyancy flux found in the experiment are compatible with the mechanism for mass transport identified in this paper.

The authors give the buoyancy flux per unit mass, normalized by $(r_i \Omega)^3/\Delta r$ and denoted $\hat{F}_e$. This quantity is related to our definition of $\mathcal{B}$ (normalized by $\mathcal{P}_{\text{lam}}$, cf. equation (3.18)) via the relation

$$\mathcal{B} = \frac{1}{4}\hat{F}_e Re(1+\eta)^2. \qquad (4.3)$$

Values of $\mathcal{B}$ for the initially linearly stratified experiments are taken from their figure 5 and reported here in table 2. These values can be compared to the results of our simulations in table 1. Whereas our buoyancy fluxes at $Sc = 1$ are never lower than 1 for $Re \geqslant 5000$, those of Oglethorpe *et al.* (2013) at larger $Re$ and $Sc$ systematically are: another indication that small scales are not helping with the buoyancy flux. Recall that $\mathcal{B}$ is a simple function of kinetic energy production $\mathcal{P}$ and mixing efficiency, i.e. $\mathcal{B} = Ri_f \mathcal{P}$. Compared with the DNS, we expect slightly larger $\mathcal{P}$ in the experiments, which are done at larger $Re$, but we also expect a significant decrease in $Ri_f$ from the large Schmidt number $Sc = 700$. Therefore, finding lower $\mathcal{B}$ in Oglethorpe *et al.* (2013)'s experiments is not incompatible with the mechanism of mass transport by nonlinear waves, even though quantitative comparisons are not possible here.

Initially five-layer stratification profiles apparently lead to similar vertical buoyancy flux across the container, leading to a universal flux law. In ??, the authors go one step further and concentrate on a two-layer stratification, in order to pinpoint the mechanism of mass transport across an isolated interface. The authors describe a periodic mixing event involving a coherent spiral structure, dominated by an $m = 1$ mode and travelling azimuthally at an angular velocity of about $\omega/m \sim 0.2$ (??*make sure... and source?*). Figure ?? in ?? representing this structure is highly reminiscent of the three-dimensional visualization of the radial velocity and density fields in figure 6 of Pt. 1. ?? also noticed the presence of a critical layer in their coherent structure, which is also reminiscent of that found by Pt. 1 in their critical linear mode at $Sc = 700$ (figure 2(*b*) therein). These similarities between the coherent structures observed on an isolated interface in ?? and in an initially linearly stratified fluid in Pt. 1 suggest they may share a common origin. An interface is indeed a strongly, almost linearly stratified layer, in an otherwise unstratified environment. First of all we know that the angular velocity of the waves travelling in a linearly stratified environment weakly depend on the stratification: they essentially travel at the mean angular velocity of the flow (experiments: Le Bars & Le Gal (2007); DNS: Pt. 1). Moreover, we have seen in our simulations that upwards buoyancy flux through a linear stratification is entirely caused by nonlinear waves, whereas turbulence hinders upwelling. We may therefore postulate that the universal flux law is caused by a somewhat similar type of nonlinear waves present in the different cases.

| $\eta$ | $Ri$ | $Re$ | $\mathcal{B}$ |
|---|---|---|---|
| | 3.47 | 14915 | $0.79 \pm 0.09$ |
| 0.389 | 3.51 | 12403 | $0.64 \pm 0.06$ |
| | 6.28 | 11461 | $0.48 \pm 0.04$ |
| | 0.32 | 24219 | $0.63 \pm 0.14$ |
| 0.195 | 0.36 | 23184 | $0.59 \pm 0.15$ |
| | 0.48 | 21011 | $0.61 \pm 0.08$ |

TABLE 2. Mean value of buoyancy flux $\mathcal{B}$ ($\pm$ one standard deviation) for the different experiments of Oglethorpe *et al.* (2013) started with a linear density profile.

The question is then: how can helical structures be found on a thin interface? We propose that the axially aperiodic structures dominating on an interface derive continuously from those found in an entirely stratified environment. Such a hypothesis could be tested by imposing Dirichlet boundary conditions on the density field such that the linear stratification term (2.4) would be suppressed at the walls, over some extent of the domain. A parameter $\alpha = l/h$ would then characterise the ratio between the depth $l$ of a linearly stratified zone to the size $h$ of the domain along $z$. The linear stratification would correspond to the limit $\alpha = 1$, and modelling an interface would then correspond to choosing a small value for $\alpha$. By homotopy between these two cases, it is plausible that helical-like patterns confined within the interface may be found. In fact, this is exactly what was found by Leclercq *et al.* (2016), which introduced localized stratified layers in a quasi-Keplerian configuration with end-plates. The unstable modes developing in the stratified layers looked very similar to their counterpart in the fully stratified system (figure 5($b$, $c$) in their article). And the spatiotemporal diagram of the experiment in their figure 10($a$) shows what appears to be a ribbon structure, created by the interaction of 'localized helical modes' in the stratified layers. The localized ribbon is responsible for upwelling in this case, as evident from the destruction of the density profile after 40 h in their figure 5($b$). Of course, interfaces are much thinner (Oglethorpe *et al.* (2013) estimated that $l \approx 1$ cm), but the local value of the Richardson number would also, as a result, be much larger. This would lead to highly 'compressed' structures, potentially able to fit within the interface.

Pursuing this idea further, we revisited the hypothesis of a strongly stratified turbulence regime made by Oglethorpe *et al.* (2013) by estimating the buoyancy Reynolds number in their experiments, using expression (4.2). The values of $\eta$, $Re$ and $Ri$ for the initially linearly stratified experiments are reported in table 2. Estimates for the dissipation can be obtained thanks to the recent contribution of Brauckmann & Eckhardt (2013), assuming stratification still has a weak impact on the torque at $Sc = 700$. We find that $\epsilon < \mathcal{P} < 20$ for experiments at $\eta = 0.389$ (where $Re_S \lesssim 2 \times 10^4$) and $\epsilon < \mathcal{P} < 30$ for $\eta = 0.195$ (where $Re_S \lesssim 4 \times 10^4$). With these estimates, we arrive at respectively $Re_b < 5$ and $Re_b < 15$ for the two configurations, both values being well-below the commonly accepted threshold of $20 - 30$ for stratified turbulence. If the estimates of dissipation are correct, then it is unlikely that the experiments were done in the regime of strongly stratified turbulence. A generic mechanism for mass transport based on nonlinear waves would, therefore, not seem incompatible.

## 5. Conclusions

This paper is the second part of a study of the dynamics of mixing in stratified Taylor–Couette flow (STC). Whereas Pt. 1 was focussed on the identification of coherent vortical structures leading to layer formation, the present paper investigates the energetics of mixing. The main conclusion is that the coherent structures, made of a superposition of nonlinear waves, are responsible for the upwelling of mass.

We investigated the energetics of the flow by deriving kinetic and potential energy budgets for our vertically periodic model of STC. Using the framework set out by Winters *et al.* (1995), we split potential energy into available and background reservoirs, in order to focus on mixing.

The kinetic energy budget revealed that stratification inhibits dissipation. Indeed, the production of kinetic energy due to the viscous torque on the inner cylinder appears to be weakly dependent on the stratification. But since a portion of the injected energy is used for upwelling, the amount of dissipation in the stratified flow must be lower than in the unstratified case. This partly explains why our simulations are only weakly turbulent, despite the large Reynolds numbers, reaching $Re = 10^4$. At larger $Sc$ number, turbulence is more vigorous: dissipation increases and the buoyancy flux decreases.

Examining the potential $E^p$ and background potential energy $E^b$ budgets lead to the observation that not all quantities are physically relevant in a periodic model. Indeed, to avoid the problem of infinite energies, we integrated all terms over one (or more) axial periods. For a quantity to be physically meaningful, it must give the same result under any arbitrary translation of the periodic part of the flow along $z$ (the governing equations are invariant under such transformations). But this is only true for the buoyancy flux, the dyapicnal mixing term and the available potential energy $E^a = E^p - E^b$. All the other quantities ($E^p$, $E^b$, and flux terms at the open axial boundaries) vary under arbitrary translations, hence their time evolution is not directly related to mixing.

After observing that the amount of available potential energy was quite small in all flows A–G, we took the limit of vanishing Thorpe scale and analysed the consequence on the potential and background potential energy budgets. We showed that the buoyancy flux and dyapicnal mixing terms are equal under that assumption. As a consequence, the flux Richardson number $Ri_f$, defined as the buoyancy flux divided by the production of kinetic energy, is an appropriate measure of mixing efficiency. The analysis of $Ri_f$ revealed a simple trend: the more disordered the flow, the lower $Ri_f$.

In order to identify the spatial scales responsible for upwelling, we took advantage of the axial periodicity of the flow and considered the (time-averaged) spatial spectrum of buoyancy flux. Strikingly, we found that the dominant positive contributions always come from the large coherent structures identified in Pt. 1, i.e. (mixed-)ribbons or (mixed-)cross-spirals corresponding to a superposition of saturated helical waves of opposite 'handedness'. Conversely, the small spatial scales of turbulence contribute weakly, and in general negatively to the buoyancy flux, confirming that disorder hinders mixing.

Going one step further, we defined a power factor associated with each spatial Fourier mode, by analogy with dipoles in alternating current. This power factor varies between -1 and 1 and characterises the correlation between the perturbation density and vertical velocity components of each mode, averaged over time and radius. In the case of nonlinear waves, the power factor characterises the phase shift between the two oscillating signals. Whereas this coupling is absent for linear inviscid waves (the classical framework for the study of the propagation of internal waves), it is the dominant mechanism for mass transport here. For $Sc = 1$, the coupling occurs through the diffusive term in the advection-diffusion equation. This linear mechanism leads to values of $Ri_f \sim 0.1 - 0.3$ at low $Sc$, which are surprisingly large given the low $Re_b = O(1)$. For larger Schmidt numbers, the diffusion term becomes weak, naturally leading to a decrease in $Ri_f$. However, a nonlinear mechanism takes over, which is also able to induce a positive correlation between the density and vertical velocity components of the most energetic modes. The combination of these two mechanisms ensures that the coherent structure is always transporting mass upwards, regardless of the flow regime. The nonlinear mechanism appears, however, to be far less efficient at producing this coupling, explaining in turn the much lower value of $Ri_f$ at larger $Sc$.

This phenomenon of mass transport by nonlinear waves is the signature of chaotic advection rather than turbulent mixing. Ribbon-type structures are indeed known to produce complex 'flow skeletons', with stagnation points in the rotating frame, from which emerge entangled stable and unstable manifolds. As a consequence, the flux Richardson number can be very well approximated by considering the coherent structure in isolation. In our work, this was achieved by following the nonlinear branch as it bifurcates supercritically from the base flow, until $Ri_f$ reaches an asymptotic value (or until the branch becomes unstable). But in general, this could be done in a more robust fashion by directly converging the coherent structure using a Newton-type solver (Wedin & Kerswell 2004), whether it is stable or not.

Next, we came back to our potential energy budgets and discussed the role of end-plates in the energetics of mixing. Two effects combine and lead to a large increase in both potential and background potential energies in practice. One of them is related to the impermeability condition, which stops the buoyancy flux and leads to accumulation/depletion of mass at the top/bottom ends of the domain. The other one is related to the no-flux boundary conditions on the density field in the case of salt stratification. This condition also leads to the growth of unstratified layers near the end-plates, by diffusion of the inhomogeneous density field. As potential energies are integrated over the volume, these unwanted effects may end up being the dominant sources of energy transfer. A clear advantage of considering energetics in an open control volume, far from the end-plates, is that it truly captures the mechanisms of mixing in the bulk of the flow.

Finally, we revisited Oglethorpe *et al.* (2013)'s experiments and found that the buoyancy Reynolds number $Re_b$ may have been insufficiently large to reach the regime of statified turbulence, despite large Reynolds numbers $Re = O(10^4)$. The presence of powerful non-axisymmetric waves in the flow, confirmed in the recent experiments of Pt. 1, is consistent with this idea. Moreover, the measurements of the buoyancy flux made by Oglethorpe *et al.* (2013) do not seem incompatible with the numerical results of this paper, even though we must exercise caution with quantitative comparisons because of the $Sc$ number mismatch. In any case, our results seem to indicate that vigorous turbulence is unlikely to lead to enhanced mass transport, as small scales were systematically found to cause downwelling, even at $Sc = 16$ and $Re_b = (10^2)$. With this in mind, we also propose that the universal flux law identified by Oglethorpe *et al.* (2013) could be caused by a similar mechanism to that identified in this paper, involving coherent structures trapped within density interfaces.

Because of the strong impact of the Schmidt number on the values of $Ri_f$, future numerical work will be targeted at large $Sc$ simulations, using high resolution DNS to check whether the mass transport mechanism uncovered in this study is still relevant at $Sc \gg 1$. If turbulence keeps transporting mass down, then coherent structures observed in experiments should remain fully responsible for the upwards buoyancy flux. From an experimental viewpoint, torque measurements would be extremely valuable, as they would not only give access to the flux Richardson number, but they would also provide an undisputable upper bound for dissipation, hence for the buoyancy Reynolds number.

Finally, the connection between initially linear and initially two-layer stratifications will be investigated further, in order to assess whether mass transport by nonlinear waves may be the key to the universal flux law discovered by Oglethorpe *et al.* (2013).

## Acknowledgements

We are indebted to Liang Shi, Markus Rampp, José-Manuel Lopez, Björn Hof and Marc Avila for sharing with us their excellent DNS code (Shi *et al.* 2015). C. L. gratefully acknowledges fruitful collaboration with R. R. Kerswell. Chantal Staquet, Jean-Marc Chomaz and Tristan Leclercq are also thanked for enlightening discussions. This work has been supported by the EPSRC, under grant EP/K034529/1.